\documentclass[preprint]{seismica}

\usepackage{xcolor}
\usepackage{soul}

\usepackage{{booktabs}}

\title{PyOcto: A high-throughput seismic phase associator}
\shorttitle{PyOcto: A high-throughput seismic phase associator} %

\author[1]{Jannes Münchmeyer
	\orcid{0000-0002-4006-9673}
	\thanks{Corresponding author: munchmej@univ-grenoble-alpes.fr}
}
\affil[1]{Univ. Grenoble Alpes, Univ. Savoie Mont Blanc, CNRS, IRD, Univ. Gustave Eiffel, ISTerre, Grenoble, France}

\begin{document}
	
\makeseistitle{
\begin{summary}{Abstract}
Seismic phase association is an essential task for characterising seismicity: given a collection of phase picks, identify all seismic events in the data.
In recent years, machine learning pickers have lead to a rapid growth in the number of seismic phase picks.
Even though new associators have been suggested, these suffer from long runtimes and sensitivity issues when faced with dense seismic sequences.
Here we introduce PyOcto, a novel phase associator tackling these issues.
PyOcto uses 4D space-time partitioning and can employ homogeneous and 1D velocity models.
We benchmark PyOcto against popular state of the art associators on two synthetic scenarios and a real, dense aftershock sequence.
PyOcto consistently achieves detection sensitivities on par or above current algorithms.
Furthermore, its runtime is consistently at least 10 times lower, with many scenarios reaching speedup factors above 50.
On the challenging 2014 Iquique earthquake sequence, PyOcto achieves excellent detection capability while maintaining a speedup factor of at least 70 against the other models.
PyOcto is available as an open source tool for Python on Github and through PyPI.
\end{summary}
}  %
	
\section{Introduction}

One of the fundamental tasks in seismology is creating detailed seismicity catalogs.
Highly complete catalogs can reveal, for example, spatial migrations, locking patterns, or changes in seismicity rate \citep{gonzalez2023relation,moutote2023evidence,tan2021machine}.
The standard workflow for event detection consists of two steps: phase picking and phase association.
The phase picking step identifies the times of seismic phases arrivals in continuous waveforms.
The phase association step aims to find consistent sets of picks that can be associated to a seismic source, called an event.
This grouping enables downstream analysis steps requiring multi-station data, for example, localisation or magnitude estimation.
In addition, phase association allows to discard spurious picks.

Traditional phase association algorithms often rely on greedy, combinatorical strategies \citep{johnson1995earthworm}.
However, these approaches scale poorly with an increasing number of picks.
While this has already become a challenge due to the growing number of seismic stations in large-scale deployments, the problem has been supercharged with the advent of highly sensitive, deep-learning-based seismic phase pickers.
Deep-learning-based pickers employ neural network models and are trained on millions of manually labeled seismic phase pick.
They outperform traditional picking models substantially in terms of sensitivity and pick precision \citep{zhu2019phasenet,mousavi2020earthquake,munchmeyer2022picker}.

To deal with this flood of phase picks, in recent years, a wave of new phase association algorithms have been published.
These approaches range from improved grid-search strategies to complex deep learning architectures.
We review the main contributions in the subsequent background chapter.
However, before we want to discuss the main challenges and performance indicators for seismic phase associators.

The key metric for seismic phase associators is the quality at which they recover seismic events.
This includes two aspects: the fraction of events being recovered, i.e. true positive rate or recall, and the fraction of identified events being incorrect, i.e. false positive rate.
Usually a tuning parameter can be used to trade-off between those metrics: either a higher recall with a higher rate of false positives or a lower recall with lower false positive rate.
The second metric concerns the same questions on pick level: how many picks have been correctly associated and how many picks have incorrectly been associated.
Similar trade-offs to the event metrics exist.
As ground-truth catalogs for seismicity are not available, seismic phase associators are usually evaluated on synthetic data, i.e., phase picks predicted using travel time calculation and random noise picks.
In addition, models are tested qualitatively on real-world example scenarios without ground-truth.

A metric often disregarded is the run time of the algorithms.
However, given the ever-growing number of picks, we consider this metric essential to understand the scalability of current algorithms and their applicability to large scale deployments.
Run time issues make some of the current associators non-applicable to such deployments, as we show in our examples where some associators did not complete associating a single day of phase picks within 48 hours.

While the recently published associators improve on all of these metrics when faced with large collections of seismic picks, our experiments show that associators are still a limiting factor when building seismicity catalogs. 
This refers to both the precision and recall of events and picks, and the run times, with several associators requiring much more time for association than the phase pickers for picking.
For this reasons, we propose PyOcto, a novel \textbf{Py}thon-based associator inspired by the \textbf{Octo}tree data structure. 
PyOcto is based on the idea of dividing space-time into potential origins.
It achieves fast run times by only exploring promising origin regions, making it a high-throughout phase associator.
PyOcto is available as an open source code with a range of different input and output interfaces for easy use.

\section{Background}

Before describing the PyOcto architecture, we introduce the most popular novel seismic phase association methods published within the last years.
All described algorithms rely on first arriving P and S phase picks without taking into account later phases.
REAL \citep{zhang2019rapid} is an optimized grid-search algorithm.
Instead of searching a full space-time grid, REAL is based on the assumption that a station close to the event will record the first P pick.
Starting with one P pick, a grid search is performed in a volume around the picking stations.
This reduces the search space from the whole study area to a smaller volume.
In addition, it removes the time dimension from the search, as the approximate origin time for each origin can be inferred from the starting pick.
REAL can use homogeneous and 1D velocity models.

HEX \citep{woollam2020hex} is a hyperbolic phase associator.
Assuming a homogeneous velocity model, it postulates that the picks of one event need to occur on a hyperbola.
HEX uses the probabilistic RANSAC algorithm to fit such hyperbolas to the picks.
In this algorithm, random candidate sets of picks are drawn and a hyperbola is fit.
If the hyperbola contains sufficiently many picks, an event is declared.

GaMMA \citep{zhu2022earthquake} is based on a similar assumption of a hyperbolic moveout but uses a different optimisation scheme.
The method interprets the picks as a Gaussian mixture with each event a different mixture component.
GaMMA uses an expectation-maximization (EM) algorithm for optimizing the clusters.
As run times for the EM algorithm grow substantially superlinearly with the number of picks, GaMMA uses DBSCAN \citep{ester1996density} to group picks before applying the EM algorithm to each cluster.
\cite{ross2023neural} proposed Neuma, a generalisation of GaMMA using an Eikonet \citep{smith2020eikonet} to enable arbitrary 3D velocity models instead of the homogeneous velocity model.

In addition to these optimization based algorithms, several deep learning models have been proposed for phase association.
PhaseLink \citep{ross2019phaselink} uses a recurrent neural network applied to pick times, phase type and station locations to identify pairwise associations between picks.
It then employs an aggregation step to infer consensus sets of matching phases that correspond to event detections.
GENIE \citep{mcbrearty2023earthquake} uses a Graph Neural Network.
Similar to PhaseLink, GENIE uses the arrival time, phase type and station location as inputs.
In contrast to PhaseLink, GENIE treats all picks jointly and outputs the full association result from the neural network.
Both GENIE and PhaseLink are trained on synthetic data generated using 1D velocity models.
The training step needs to be conducted once for each target region, afterwards the models can be applied to arbitrary amounts of data. 

\section{Methods}

\begin{figure}[ht!]
\centering
\includegraphics[width=0.5\textwidth]{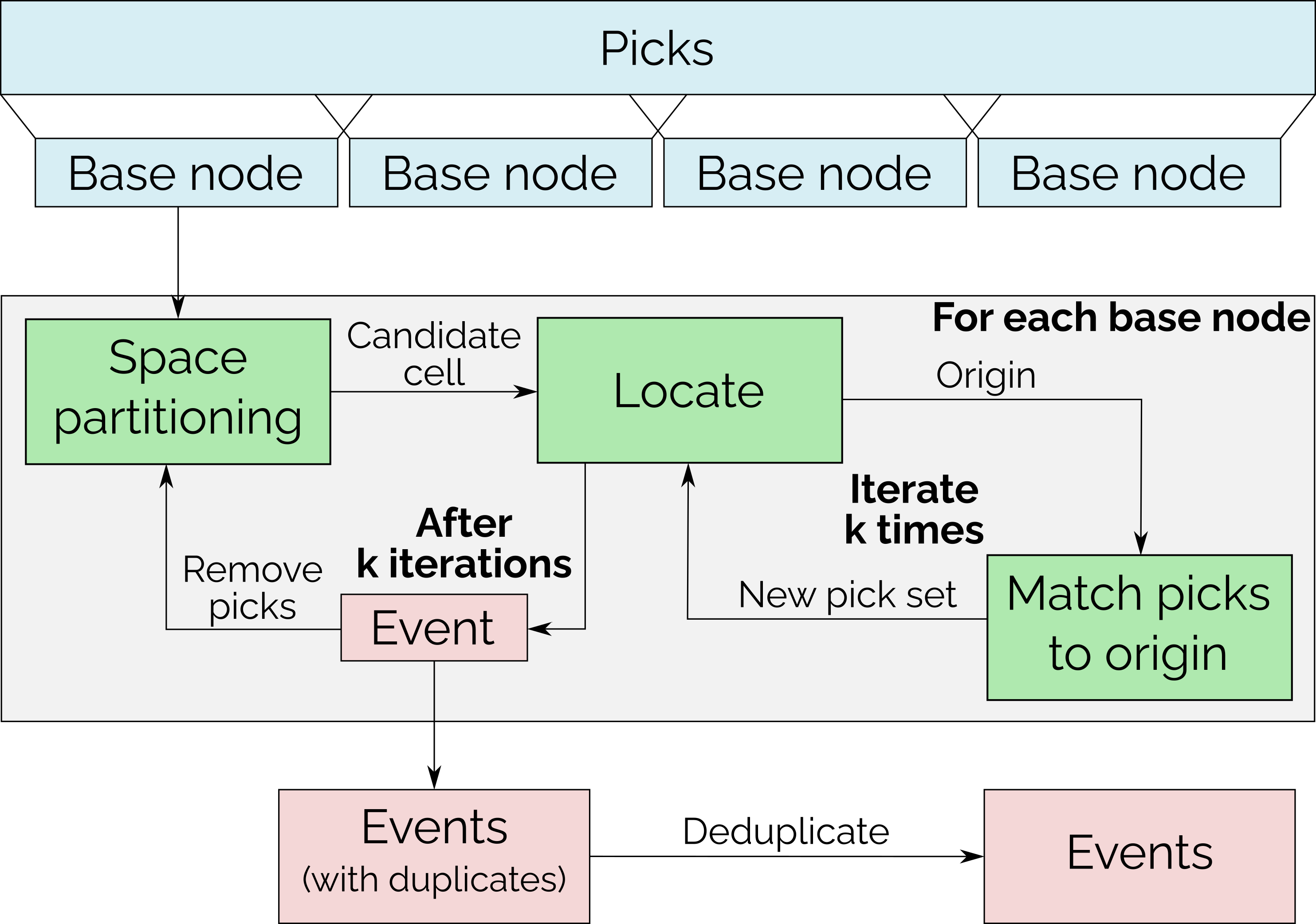}
\caption{Schematic view of the full PyOcto pipeline. The picks are split by time into base nodes. For each base node, the grey box is executed. Several of these boxes can be executed in parallel. Within each box, the space partitioning algorithm (see Figure~\ref{fig:octotree_schematic} and the localisation/pick matching steps are conducted. Events are output and finally deduplicated.}
\label{fig:scheme}
\end{figure}

In the following sections we present the PyOcto associator.
We start with the core algorithms and then discuss details, optimisations and implementation details.
A schematic overview of the full associator is provided in Figure~\ref{fig:scheme}.
Throughout the description we add the parameter names used in the implementation in italics in brackets to allow easier cross-referencing.

\subsection{Core algorithm}

\begin{figure*}[ht!]
\centering
\includegraphics[width=\textwidth]{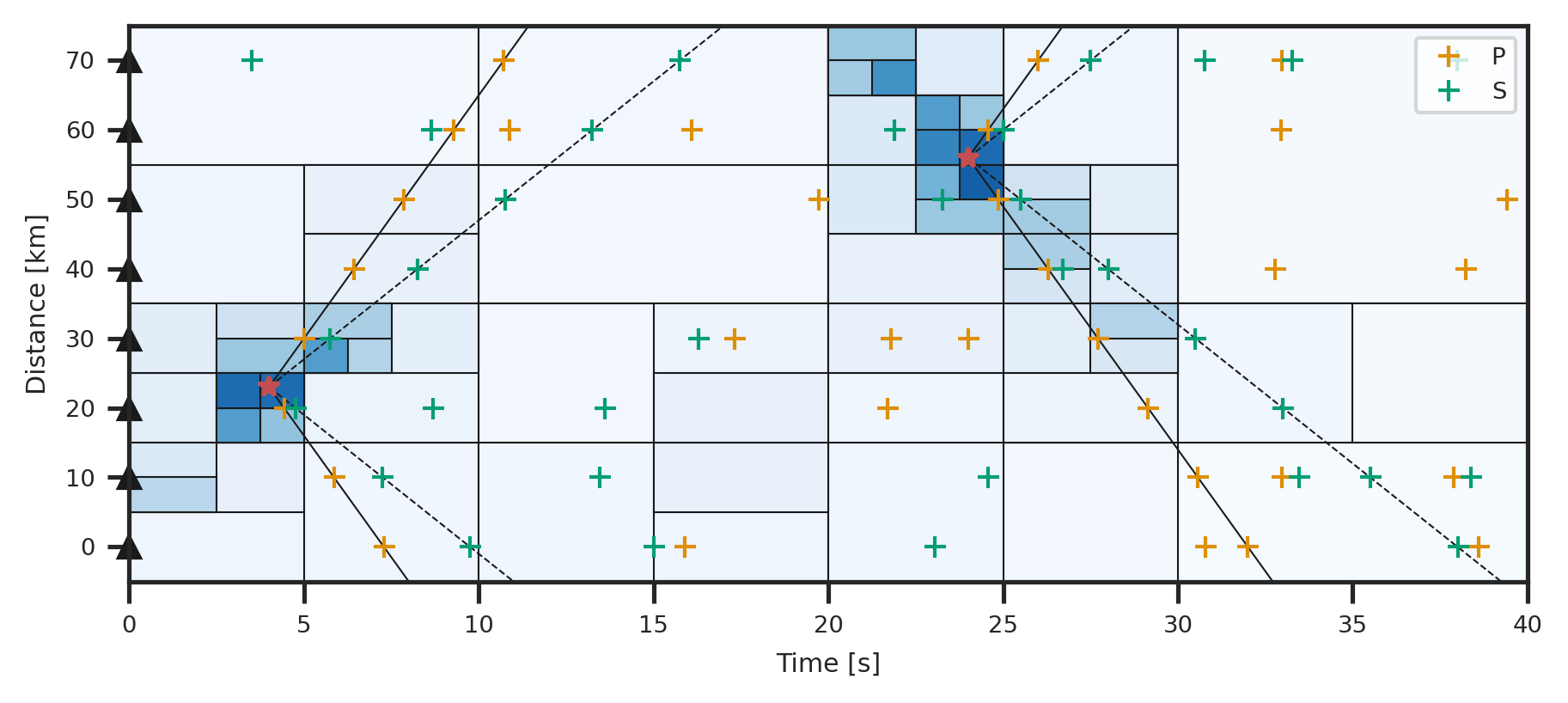}
\caption{Schematic view of the gridding scheme with only one spatial dimension and the time dimension. Picks are indicated by crosses, the station locations are marked on the left by black triangles. Two events are contained, marked by red stars with P (solid) and S (dashed) moveout shown in black. The background shows the gridding with each cells shading corresponding to the number of picks per area . Only cells with at least 6 matching P picks and 6 matching S picks were explored. For each area, only the smallest cell explored is shown, i.e., all larger cells explored before in the same region are not visualised.}
\label{fig:octotree_schematic}
\end{figure*}

PyOcto is based on partitioning space-time into cells.
The key idea is to mimic a grid-search associator while only looking at ``useful'' grid cells.
We achieve this by using a data structure inspired by an octotree with an additional time axis.
The data structure consists of a collection of 4D volumes (3D in space, 1D in time), that we will call nodes in the following to highlight the resemblance of a tree data structure.
We show a simplified version of this with only one space axis in Figure~\ref{fig:octotree_schematic}.
Each volume/node $V$ is associated to the list of picks $picks(V)$ that could have originated from the node.
More formally, let $V$ be a node and $(s, t)$ a pick at station $s$ at time $t$.\footnote{For simplicity we omit the phase of the pick here. The inclusion of phase type is natural and only involves taking different travel time models for P and S waves.}
We write
\begin{align}
(s, t) \in picks(V) \Leftrightarrow \exists (x_0, t_0) \in V : t_0 + tt(x_0, s) = t + \epsilon
\label{al:invariant}
\end{align}
with $tt(x_0, s)$ the travel time from the origin $x_0$ to the station $s$.
We include an $\epsilon$ to indicate that the equation only needs to hold up to a given uncertainty (\textit{tolerance}).
This uncertainty takes into account inaccuracies in the velocity model and the pick times.

There are two crucial insights about the definition of picks belonging to a node.
First, while for each pick there exists a location/time in the node where it could have originated, this location/time might be different for each pick.
Therefore, a set of picks originating from a node is not a sufficient condition for associating these picks into an event.
This becomes obvious when looking at very large nodes.
On the other hand, it is a necessary condition, i.e., if there is an event with sufficiently many picks in the dataset, there must be a node that contains all these picks.
Second, the assignment of picks to nodes is not unique.
A pick might be contained in multiple nodes, even if these nodes are disjoint.
However, only few nodes will contain enough picks to produce an event.
The key idea of PyOcto is to cleverly identify these nodes.

PyOcto starts with a large node spanning the whole study area and a long time.
All picks recorded during this time (with adjustments for boundary effects) can be assigned to the node.
We initialize a list of active nodes with this node.
The association then repeatedly takes the active node with the highest number of picks and performs one of the following actions:
\begin{itemize}
\item if the node can not create an event anymore: discard node
\item if the node is small enough: try creating an event
\item otherwise: split the node and add children to the list of active nodes
\end{itemize}
We use a priority queue for the list of active nodes to efficiently retrieve the node with the highest number of picks.
In the following, we describe the different actions.

\textbf{Splitting a node: }
The most common action is splitting a node.
For this action, we split the node $V$ into two disjoint children $V_1$ and $V_2$, such that $V = V_1 \cup V_2$.
We split $V$ in half along the coordinate axis in which $V$ has the largest extent.
To compare the time axis, we multiply it with a constant velocity, by default 5~km/s.

We then build the sets $picks(V_1)$ and $picks(V_2)$ by iterating over all candidates in $picks(V)$.
This check can easily be performed using equation~(\ref{al:invariant}).
As noted before a pick can be assigned to both of these sets at the same time.

\textbf{Discarding a node: }
Essential for the high performance of PyOcto is to discard nodes early if they can not produce an event anymore.
For this, we use the following criteria:
\begin{itemize}
 \item minimum number of total picks (\textit{n\_picks})
 \item minimum number of P picks (\textit{n\_p\_picks})
 \item minimum number of S picks (\textit{n\_s\_picks})
 \item minimum number of stations with both P and S picks (\textit{n\_p\_and\_s\_picks})
\end{itemize}
All thresholds are configurable and should be adjusted to the dataset.
As a subvolume can never contain more picks than the parent node, once a node violates any of these criteria it can not create an event anymore and can be discarded.

\textbf{Creating an event: }
If a cell is smaller than a predefined threshold along all axes (\textit{min\_node\_size}), PyOcto tries creating an event.
For this, we locate an event based on all picks in a cell.
The full localisation procedure is described in \ref{sec:localisation}.
We then identify whether all picks fit the determined location and remove potential outliers.
These outliers might occur as not all picks in the node need to necessarily stem from the same source location/time.
In addition, we scan all other picks to identify if further picks are consistent with the list of picks.
This operation can be performed efficiently using a binary search in time.
We add these picks to the list of picks.
This procedure is repeated multiple times (\textit{refinement\_iterations}), by default 3, to stablise the event.
If at any point in this iteration the picks do not fulfill the conditions for nodes outlined above, the event creation is stopped as unsuccessful.

Even though the node already gives a preliminary location and station set, the location procedure is required for multiple reasons.
First, while the node groups a candidate set of picks, there is no guarantee that all of these can be associated to a common origin.
Second, the optimal location for a set of picks does not necessarily need to fall within the node, in particular, because the same set of picks can be contained in multiple nodes.
This is also the reason why it might be possible to associate additional picks to the location.
While traversing the nodes by number of picks makes it likely to select nodes already containing the majority of picks for an event, this can not be guaranteed in face of spurious picks.

In contrast to some other associators (e.g., GaMMA), PyOcto can not use amplitude information for association.
However, obtaining accurate amplitudes for events at low signal-to-noise leves, as for the majority of events detected with deep learning, is challenging.
From our anecdotal experiments on real data, we did not see a major advantage from the use of amplitude information.

\subsection{Localisation procedure}
\label{sec:localisation}

To identify the most likely origin for a set of picks, we use the equal differential-time (EDT) loss \citep{lomax2000probabilistic}.
Compared to an L2 loss on the travel time residual, the EDT loss has two advantages.
First, it is independent of the origin time, thereby reducing the search space.
Second, it is more stable against outlier picks.
As we expect outliers to be contained in our pick set, this is a useful property for our application.

To find the minimum of the EDT loss, we use a greedy algorithm.
Starting with the whole study volume, we split the volume in half $k$ times (\textit{location\_split\_depth}) into $2^k$ subvolumes.
For each subvolume, we calculate the EDT loss at the volume center.
From the volume with the lowest EDT loss, we go up $l$ splits (\textit{location\_split\_return}).
This volume, with a size of $2^{k - l}$ is used as the new start for the search and we repeat the splitting and search procedure.
We iterate this step until the volume reaches a predefined size (\textit{min\_node\_size\_location}).

This greedy algorithm has a trade-off between accuracy and runtime.
When splitting the volume into only few pieces and only using a low $l$, this leads to low run-time but potentially suboptimal minima.
On the other hand, too fine splitting in each step will increase runtimes at virtually no gains in location accuracy.
We set the default to $k=6$ and $l=4$, but make the parameter individually configurable.
We note that insufficient values for $k$ and $l$ can lead to striping artifacts, i.e., locations at the edges of larger volumes caused by insufficient sampling.

\subsection{Velocity models}

At its core, PyOcto relies on travel times.
These travel times need to be obtained from seismic velocity models.
Two types of queries occur in the PyOcto algorithm.
First and most commonly, volume queries of type $(s, t) \in picks(V)$, i.e., identifying if a pick can originate from a volume.
Second, for the localisation algorithm, traditional travel times between the proposed origin and the station are required.
Both of these queries will be executed in very high numbers and therefore need to be implemented efficiently.

PyOcto implements two velocity models, a homogeneous model and a 1D layered model.
For the homogeneous model, we assume constant P and S velocities.
To solve the volume query, we identify the earliest and latest times a pick from the volume could arrive at the station.
The earliest time is achieved by the earliest origin time in the volume plus the travel time to the closest point in the volume.
Similarly the latest time can be derived using the point with the highest distance to the station.
The derivation of the travel times from a fixed origin are trivial using Pythagoras theorem.
Both queries run in $\mathcal{O}(1)$ time.

For the layered velocity model, we use a precalculation step to substantially improve performance.
First, we calculate P and S arrival times on a dense grid using an eikonal solver.
This step takes a few seconds but only needs to be run once.
For extracting travel times we run 2D bilinear interpolation between the 4 closest grid nodes.
For the area queries, i.e., if a pick can result from a volume we use the observation that for a 1D velocity model, the shortest travel time must be at the closest epicentral distance and the longest travel time at the furthest.
However, it is not a priori clear at which depth these times occur.
Potential candidates are the shallowest and deepest points of the queried depth interval, plus all local extrema within the depth interval.
To efficiently query the local extrema, we cache all local extrema at each distance.
As for typical velocity models each distance has at most a handful of local extrema, they can simply all be checked when necessary.
In addition, to correct for station elevation, we add an elevation correction based on a constant velocity and vertical incidence.
While this is an approximation, errors are negligble for association purposes.
To determine travel times for localization, we use bilinear interpolation between the 4 closest precalculated travel times.

PyOcto does not support 3D velocity models as performing efficient, i.e., constant run time, volume queries as required for the splitting algorithm is non-trivial.
This is identical to most common algorithms, that are limited to homogeneous or 1D models.
In contrast, deep learning models are able to use arbitrarily complex models.
PyOcto supports different velocity models for the splitting and the localisation step.
In principle, it would be easy to extend the localisation step to 3D models.
However, we have not tested this and only expect substantial improvements for regions with velocity structures strongly deviating from a layered model.

PyOcto supports station terms, i.e., constant time offsets for phase arrivals at a station, which can occur due to local structure.
We implement additive station terms, i.e., the station term is added to the predicted travel time from the velocity model.
This is the same sign convention as used by NonLinLoc \citep{lomax2000probabilistic}.
Station terms are not determined dynamically but have to be defined before running the association.
However, they can be obtained by iteratively running PyOcto and inferring station terms from the residuals of the previous run.

For efficient calculation of distances, PyOcto relies on local coordinate transforms.
By default, we suggest transverse Mercator projections.
The transformation from latitude and longitude values to local coordinates needs to be performed only once before the association step.
While distance measures will become inaccurate for very large study areas, we did not observe any issues in our case studies with diameters up to $\sim1500$~km.

\subsection{Initialisation}

As described in the introduction of the algorithm, the association starts with a node spanning the whole study area.
In principle, this node could also span the whole study time.
However, in practice this is suboptimal because it will require several costly splits along the time axis that are mostly trivial.
Instead, we do not start with a single node but with a list of base nodes.

Each base node spans the whole study area but only a part of the time.
For this, we split the time into regular, non-intersecting segments, by default 20 minutes long (\textit{time\_slicing}).
Each segment is then filled with all picks that originate during the segment plus the ones occuring in a buffer time before the start of the segment (\textit{time\_before}).
This buffer time should be roughly the maximum travel time through the study area.

As two subsequent base nodes might both contain all picks for one event, the early splitting might lead to duplicate events.
For this reason, we deduplicate the events after all base nodes have been processed.

\subsection{Optimisations}

While the splitting algorithm with early stopping is a solid basis for an efficient algorithm, several points need to be taken into account that might affect runtime.
Before going into details, we review the general runtime principles.
While a formal analysis of algorithm complexity is difficult, we can make several observations.
First, run time crucially depends on the number of nodes processed.
It is therefore essential to stop the processing of each branch of the search tree as early as possible.
Second, location procedures are expensive as they require many travel time queries.
They should therefore not be triggered too often.
Based on these observations, we define multiple optimisations.

As a first optimisation, PyOcto keeps track of all picks that have already been assigned to events.
Once a pick has been assigned to an event, it is not considered anymore and removed from all nodes.
Without these picks, the adjacent nodes most likely will not fulfill the necessary minimum number of picks.
This step substantially improves runtime, as events will usually produce many adjacent nodes with high numbers of picks which do not need to be processed multiple times.

The second observation treats the case of a group of picks that can not be associated to a common origin.
The same group of picks often appears at many neighboring nodes.
As trying to create an event from these picks does not yield a consistent origin, these picks are not marked as used.
As often many neighboring cells contain the same set of picks, this leads to repeated but useless tries of locating the same set of picks.
To mitigate this situation, we cache all sets of picks that have been processed as candidate sets for localisation.
If a set has been processed before, it will be skipped in the next try.
Note that this optimisation only works because the location search depends only on the pick set but not on the location of a node.

The last optimisation is relevant in the case of a large number of stations with spurious picks.
With a growing number of stations, it becomes likely that a set of distant stations by chance produces picks that can be associated.
This does not only lead to false detections but also substantially increases run time.
At the same time, these false events are easy to identify manually because of the inconsistent pick pattern, i.e., the existence of many non-picking stations between the picking stations.

To remove this issue, we introduce two distance conditions, a relative and an absolut condition.
The absolute condition is a simple cutoff on the maximum distance between stations and sources for the space partitioning (\textit{association\_cutoff\_distance}).
This condition excludes picks too far from a given cell when checking if the pick could have originated there.
However, in the localisation and pick matching step, all picks are taken into account, ensuring that the output contains all associated picks even at larger distance.
This condition is most helpful in large, homogeneous networks and in networks without large amounts of out-of-network events.

For the case of inhomogeneous networks or networks with substantial out-of-network events, we introduce a relative distance condition, based on the assumption that it is unlikely for a station to detect an event if many closer stations did not detect it.
For every distance from a volume, we can calculate the fraction of stations within this distance that have at least one pick compared to the total number of stations.
We then identify the maximum distance where this fraction is still above a predefined threshold (\textit{min\_pick\_fraction}).
All picks at stations above this threshold are removed.
As the nodes have a spatial extent, for each station we choose the distance maximizing the number of retained picks.
This means that for stations with picks we use the minimum distance to the node while for all other stations we use the maximum distance.

While this optimization yields substantial runtime improvements for datasets with high numbers of stations, it comes at a cost.
To check the condition, at every node the distance to all existing stations needs to be calculated.
For small deployments, associations by chance are anyhow unlikely, rendering the additional runtime mostly useless.
The optimisation can therefore be deactivated.

Lastly, PyOcto uses a memory protection strategy.
As PyOcto processes nodes ordered by their number of picks, it needs to always hold a queue of active nodes.
This can, in the worst case, degrade into a breadth-first search, which is very memory intensive.
Therefore, once the total number of nodes exceeds a predefined threshold (\textit{queue\_memory\_protection\_dfs\_size}), PyOcto processes the next nodes using depth-first search.
This is highly memory efficient, as only the current call stack needs to be kept in memory.
At the same time, this can lead to increased run times.
In our experiments, this optimization was only required for very large sets of picks in short times ($\gg$100,000 picks per day).

\subsection{Implementation}

PyOcto is implemented in Python and C++.
The interface of PyOcto is implemented in Python to provide an accessible interface in a common scripting language.
Inputs and outputs are passed as Pandas data frames.
PyOcto has a slim set of dependencies.
The backend of PyOcto is implemented in C++.
The functions are natively embedded into Python using pybind11.
The association function is parallelised using pthreads.
Parallelisation is achieved by assigning base nodes to threads.
This causes very low synchronisation overhead as only the base node queue and the event list are shared between threads.
The list of used picks is not shared between threads, instead events are deduplicated at the end of the association step.
By default, PyOcto uses all available threads.
However, the thread count can be set manually (\textit{n\_threads}).

To allow an easy experimentation with PyOcto, the software implements several compatibility interfaces:
\begin{itemize}
 \item a function to read the input format from GaMMA \citep{zhu2022earthquake}
 \item a function to read the input format from REAL \citep{zhang2019rapid}
 \item a function to process SeisBench picks \citep{woollam2022seisbench}
 \item a function to use obspy Inventory objects as input \citep{beyreuther2010obspy}
 \item an output interface for NonLinLoc \citep{lomax2000probabilistic}
 \item an automated selection strategy for local coordinate transforms
\end{itemize}

PyOcto is available as open source code under MIT license, a permissive open-source license.
Pre-built wheels for Linux, Mac OS, and Windows are available on PyPI and can be installed using pip.

\section{Benchmark on synthetic catalogs}

\begin{table*}
\caption{Dataset statistics for the shallow seismicity scenario. We do not differentiate between P and S picks as both are generated in almost equal number. The picks per station include the noise picks.}
\label{tab:statistics_shallow}
\begin{tabular}{rlrrrrr}
\toprule
Events & Noise & Event picks & Noise picks & Total picks & Picks per event & Picks per station \\
\midrule
100 & 0.3 & 4,047 & 1,214 & 5,261 & 40.47 & 52.61 \\
100 & 1.0 & 4,894 & 4,894 & 9,788 & 48.94 & 97.88 \\
100 & 3.0 & 5,257 & 15,771 & 21,028 & 52.57 & 210.28 \\
500 & 0.3 & 25,658 & 7,697 & 33,355 & 51.32 & 333.55 \\
500 & 1.0 & 24,525 & 24,525 & 49,050 & 49.05 & 490.50 \\
500 & 3.0 & 23,646 & 70,938 & 94,584 & 47.29 & 945.84 \\
2,000 & 0.3 & 101,614 & 30,484 & 132,098 & 50.81 & 1320.98 \\
2,000 & 1.0 & 98,680 & 98,680 & 197,360 & 49.34 & 1973.60 \\
2,000 & 3.0 & 94,710 & 284,130 & 378,840 & 47.35 & 3788.40 \\
\bottomrule
\end{tabular}
\end{table*}

\begin{figure*}[ht!]
\centering
\includegraphics[width=\textwidth]{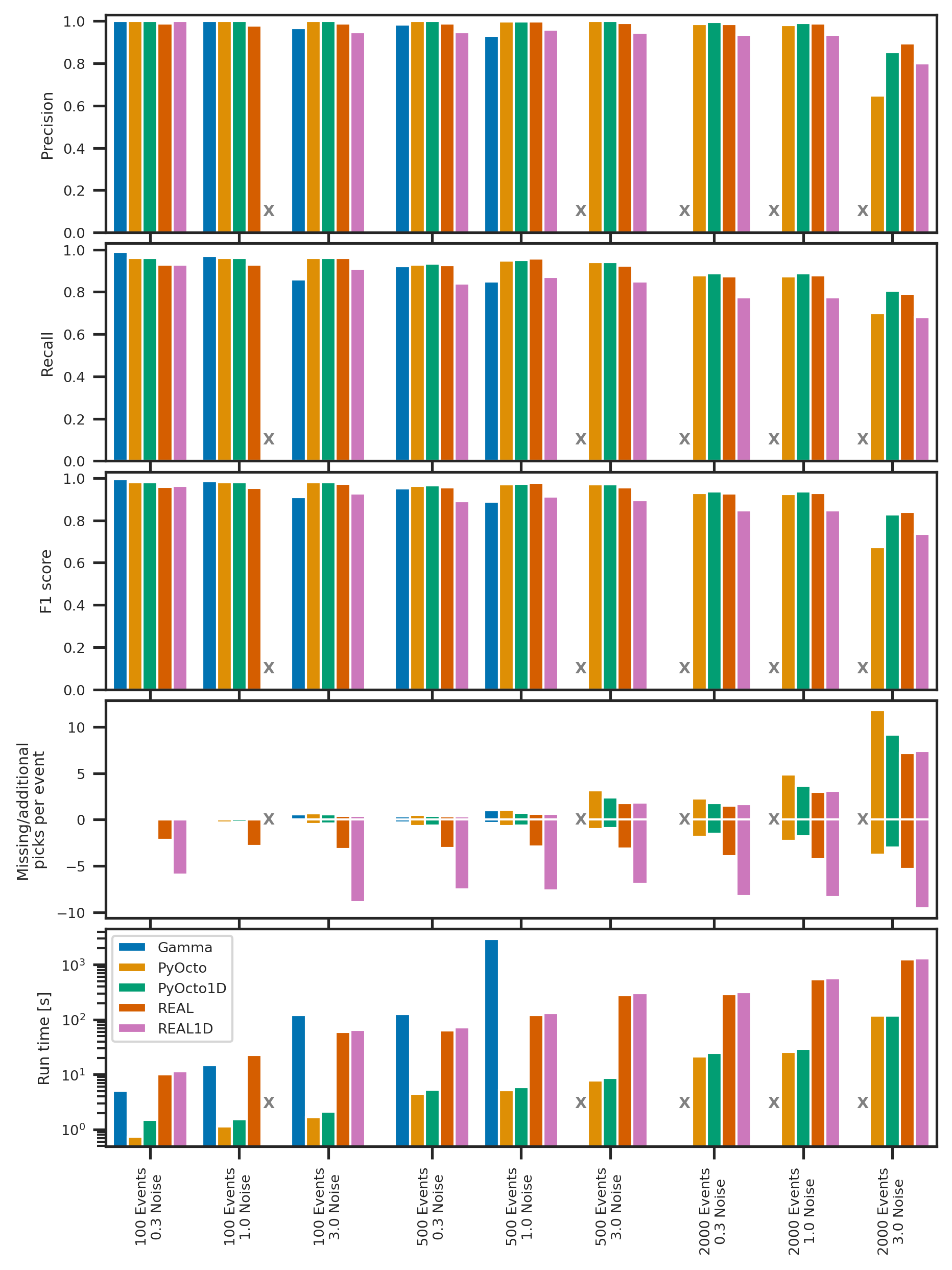}
\caption{Synthetic evaluation of the different associators in the shallow seismicity scenario. Each associator is indicated by a color. For the missing/additional picks, missing picks are indicated with a bar below 0, additional picks with a bar above 0. Missing results due to exceeded runtimes are indicated by grey Xs. A result for REAL 1D with 100 events and 1.0 noise is not available as the model reproducibly crashed with a segmentation fault. All results in numerical form are reported in Table~\ref{tab:full_benchmark}.}
\label{fig:results_shallow}
\end{figure*}

\begin{table*}
\caption{Dataset statistics for the subduction scenario. We do not differentiate between P and S picks as both are generated in almost equal number. The picks per station include the noise picks.}
\label{tab:statistics_chile}
\begin{tabular}{rlrrrrr}
\toprule
Events & Noise & Event picks & Noise picks & Total picks & Picks per event & Picks per station \\
\midrule
100 & 0.3 & 2,241 & 672 & 2,913 & 22.41 & 145.65 \\
100 & 1.0 & 2,331 & 2,331 & 4,662 & 23.31 & 233.10 \\
100 & 3.0 & 2,142 & 6,426 & 8,568 & 21.42 & 428.40 \\
500 & 0.3 & 11,414 & 3,424 & 14,838 & 22.83 & 741.90 \\
500 & 1.0 & 11,194 & 11,194 & 22,388 & 22.39 & 1119.40 \\
500 & 3.0 & 10,818 & 32,454 & 43,272 & 21.64 & 2163.60 \\
2,000 & 0.3 & 45,544 & 13,663 & 59,207 & 22.77 & 2960.35 \\
2,000 & 1.0 & 45,011 & 45,011 & 90,022 & 22.51 & 4501.10 \\
2,000 & 3.0 & 45,213 & 135,639 & 180,852 & 22.61 & 9042.60 \\
\bottomrule
\end{tabular}
\end{table*}

\begin{figure*}[ht!]
\centering
\includegraphics[width=\textwidth]{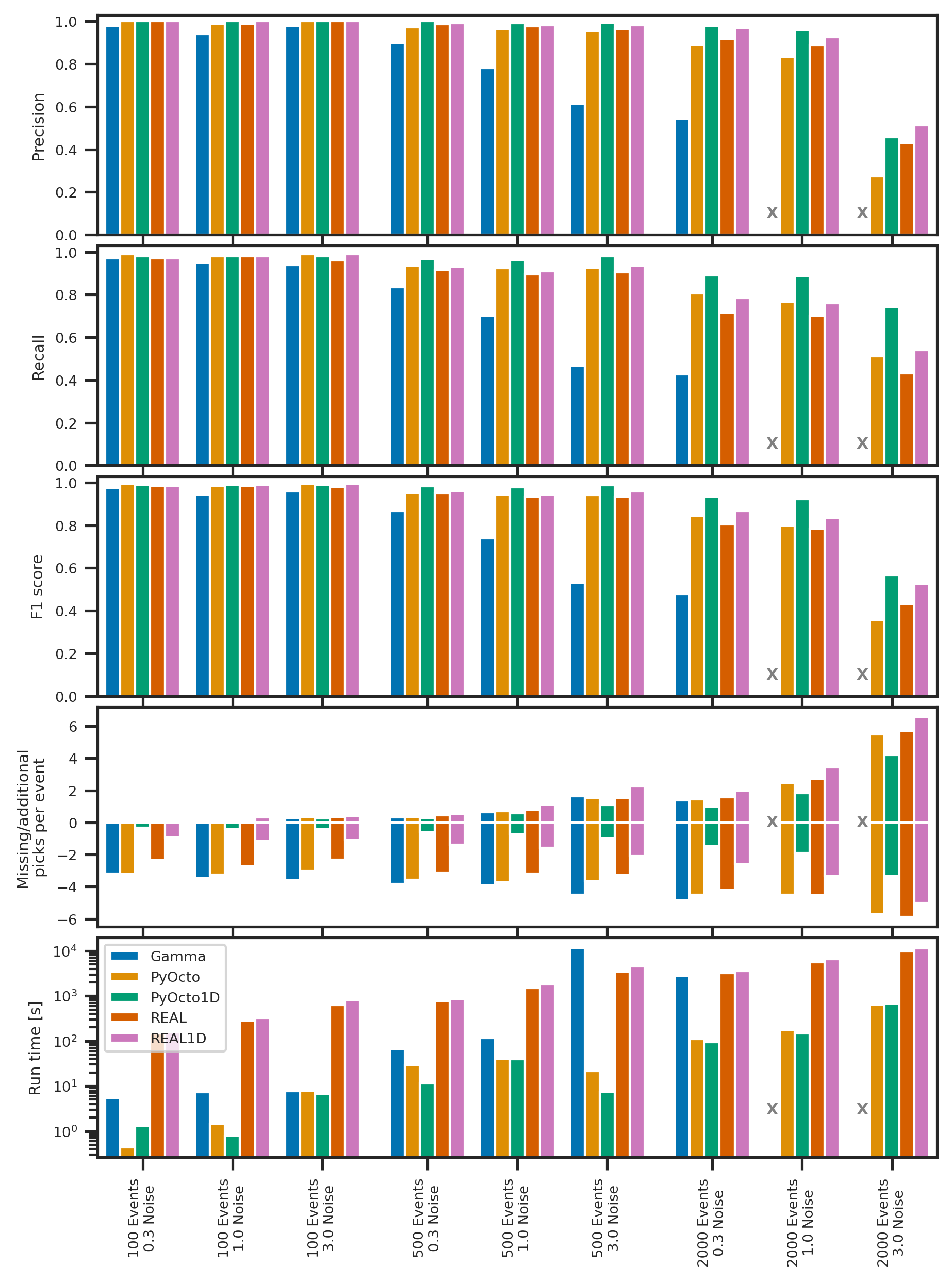}
\caption{Synthetic evaluation of the different associators in the subduction scenario. For further details see the caption of Figure~\ref{fig:results_shallow}. All results in numerical form are reported in Table~\ref{tab:full_benchmark}.}
\label{fig:results_chile}
\end{figure*}

\subsection{Setup}

To quantitatively assess the quality of PyOcto, we test it on synthetic catalogs.
We use two complementary scenarios: (i) uniformly distributed seismicity in a shallow layer; (ii) realistic subduction zone seismicity.
We compare the proposed PyOcto algorithm to two established associators: GaMMA and REAL.
We choose these algorithms as they have well-documented, open-source implementations and have both been used in numerous application cases already \citep{wilding2023magmatic,gonzalez2023relation,tan2021machine,liu2020rapid}.
We do not compare to any deep learning associators, as optimizing these pickers requires substantially more parameter choices and a fair comparison is therefore harder to guarantee.
Note that this study in not intended as a full-scale benchmark of seismic phase associators as this would be out of scope for the paper.
Instead, we restrict ourselves to this smaller-scale case study.

Both scenarios use the same procedure for data generation.
Each test case consists of one day of seismicity with a predefined number of events and a predefined noise rate.
For each event, we draw a source time uniformly within the day and draw a location and a magnitude from the distributions described below.
Based on the magnitude and hypocentral distance, we estimate detection probabilities at each station.
From these probabilities we randomly select whether a station has a P and an S arrival using correlated Bernoulli variables with correlation 0.5 between the two phases.
We predict travel-times using a 1D velocity model from \cite{graeber1999three}.
To each individual travel-time we add a Gaussian random normal variable with standard deviation 1~\% of the total travel time but at least 0.4~s standard deviation.
Finally, we add noise picks not associated to any event to the data set.
The number of noise picks is defined as the product of the number of event picks times the user-defined noise rate.
For each pick, the phase, time and station are drawn according to a uniform random distribution.
We use event numbers of 100, 500 and 2000, and noise rates of 0.3, 1.0 and 3.0. 

We compare PyOcto to GaMMA and REAL.
For each model, we manually selected reasonable parameters.
All parameters are reported in Tables~\ref{tab:params_gamma}, \ref{tab:params_pyocto},  and \ref{tab:params_real}.
For PyOcto and REAL we report results for the versions with homogeneous velocity models and 1D layered velocity models.
We provide the associators with the same velocity model we used for data generation.
We therefore expect slightly too optimistic performance results for the 1D models, however, the comparison between these models should still provide reasonable results.

For all associators we require at least 10 picks for an event detection.
We furthermore require at least 4 stations with both P and S pick for REAL and PyOcto.
We do not enforce the last condition for GaMMA as the option is not implemented.
We ensure that all events in our synthetic catalogs fulfill these conditions.

We evaluate the associators based on 6 metrics: precision, recall, F1 score, missing picks per event, incorrectly associated picks per event, and run time.
Precision is the fraction of cataloged events among all detections.
Recall is the fraction of events detected among all cataloged events.
F1 score is the harmonic mean of precision and recall.
To calculate these metrics, we define matches between cataloged and detected events through their picks.
A cataloged event $A$ and a detected event $B$ are considered a match if at least 60~\% of the picks of $A$ are also picks of $B$ and vice versa.
We use a pick-based matching instead of a location- and time-based matching as it is more stable for high event densities.

We execute the test on 16 virtual CPU cores with 8 physical cores and 64 GB main memory.
We measure runtimes from the invocation to the output of the models.
We do not measure data-independent preprocessing steps such as velocity model building as these steps only need to be executed once in an application scenario.
Exact machine configurations can vary slightly between tests, therefore the reported runtimes should be interpreted rather as an indication than an exact measure.
We limit the total aggregated runtime of all tests per associator to 48~h.
All tests not finished at this point are reported as missing.

\subsection{Uniform shallow seismicity}

As a first scenario, we study shallow seismicity.
We use 100 stations arranged in a 10x10 grid with a station spacing of $0.2^\circ \times 0.2^\circ$.
Event locations are randomly distributed within the network with a depth up to 30~km.
No out-of-network events are generated.
Magnitudes are generated from a Gutenberg-Richter distribution with a minimum magnitude of 0.5 and $b=1$.
Dataset statistics are reported in Table~\ref{tab:statistics_shallow}.

Figure~\ref{fig:results_shallow} shows the performance metrics for the shallow scenario.
Full results in numerical form can be found in Table~\ref{tab:full_benchmark}.
PyOcto and REAL obtained results for all cases with both the homogeneous and the 1D velocity model.
GaMMA did not provide solutions for the cases after 500 events and a noise factor of 1.0 as the computation did not finish within the 48~h time limit.

In all cases, PyOcto achieves the highest F1 score or a result within 0.01 F1 scode of the best model. \
The 1D model slightly outperforms the homogeneous model.
REAL with a homogeneous model achieved a slightly worse performance, followed by REAL with a 1D model.
GaMMA shows a clear degradation in F1 score with growing number of event or noise picks but still achieves good performance (F1 $\geq$ 0.89) for all cases where solutions were obtained.
For the case with 2000 events and a noise factor of 3.0, REAL (homogeneous, 0.84) performs best, closely followed by PyOcto (1D, 0.83), REAL (1D, 0.74), and PyOcto (homogeneous, 0.67).
We suspect that REAL shows slightly better performance here because the actual grid search is less affected by noise picks than the approximation using space partitioning used in PyOcto.
We note that this case is extremely challenging with each station reporting on average one pick every 23~s.

Up to 500 events and a noise rate of 1.0, PyOcto (1D and homogeneous) and GaMMA are very exact in terms of picks with few additional or missed picks.
In contrast, REAL (homogeneous) misses roughly 3 picks per event, REAL (1D) between 5 and 10.
While we are not fully certain about the missed picks, we assume it is because REAL discards picks based on the ratio between station residuals and event residuals, i.e., a low average pick residual for an event will lead to discarding picks with higher residuals even if their absolute value is not excessively high.
We note that the number of missed picks for REAL could likely be reduced through targeted parameter tuning.
For configurations with high numbers of events, in particular, in conjunction with high noise, REAL and PyOcto both include false picks with the events.
PyOcto includes more false picks than REAL, again likely related to selection criteria.
The homogeneous version of PyOcto produces about 1.5 times as many false picks as the 1D variant, likely because of the overall higher tolerance value necessary to mitigate the less accurate velocity model.

In terms of run time, PyOcto substantially outperforms GaMMA and REAL in all cases.
The run time factor between PyOcto and the next-fastest methods exceeds 10 in almost all cases, often even reaching factors of 20 and above.
Run times for the homogeneous and the 1D velocity model for PyOcto are almost identical in all cases.
We suspect that while the travel time lookup for the 1D velocity model is slightly more expensive than for the homogeneous model, this effect is offset by more focused origins from the better travel times, leading to fewer nodes that need to be explored.

\subsection{Subduction zone}

For the subduction zone scenario, we base our catalog on the IPOC network \citep{fdsn_cx} and the catalog by \cite{sippl2018seismicity}.
We chose the deployment and the catalog as a typical example of a well-instrumented, highly active subduction zone with diverse seismicity.
We draw event locations and event magnitudes independently from the catalog.
We use the IPOC stations, in total 20 stations.
The study area covers approximately $5^\circ$ North-South and $3^\circ$ East-West up to a depth of 200~km.
Out-of-network seismicity is located up to $1^\circ$ from the network.
This is a typical challenge for associators in subduction zones were offshore events will occur substantially outside the network.
Dataset statistic are reported in Table~\ref{tab:statistics_chile}.

The results in the subduction scenario largely mirror the ones from the shallow scenario but with nuanced differences that we point out in the following (Figure~\ref{fig:results_chile}, Table~\ref{tab:full_benchmark}).
First, the difference between 1D and homogeneous models is more pronounced with 1D models clearly outperforming homogeneous models in terms of F1 score.
Furthermore, the homogeneous models (GaMMA, REAL, and PyOcto) consistenly miss around 2.5 picks per event.
This highlights that the assumption of a homogeneous velocity model is insufficient for subduction zones.
Nonetheless, PyOcto and REAL with homogeneous velocity model still achieve F1 scores consistently above 0.93 for cases with up to 500 events.
In contrast, GaMMA performance clearly is below the other models already at 100 events per day and degrades substantially above.

Second, among the 1D models, PyOcto outperforms REAL more clearly than in the shallow case.
It consistently exhibits a higher F1 score and lower numbers of missed and false picks.
Even at 2000 events with a noise rate of 3.0 (with a pick per station on average every 9.5~s), PyOcto still achieves an F1 score of 0.57.

Third, run time differences are even more pronounced with PyOcto outperforming REAL often by a factor of 1000.
This is caused by the larger search grid required by REAL to handle the depth and the out-of-network events.
We note that we already reduce the impact of the larger grid size for REAL by using a larger grid spacing for the subduction scenario.
In contrast, PyOcto can easily handle large search domains due to its splitting approach that scales logarithmically with volume.
For the subduction scenario, PyOcto with a 1D model in most cases only needs about half the time of PyOcto with a homogeneous velocity model.
This suggests that the more accurate velocity model leads to fewer nodes needing to be explored.
GaMMA shows competitive run times compared to PyOcto and REAL for cases with 100 events but run times substantially exceed the ones of REAL (and thus even more PyOcto) at 500 events and above.
No solutions for 2000 events and noise rates of 1.0 and 3.0 could be obtained.

\section{Application to the 2014 Iquique sequence}

\begin{figure*}[ht!]
\centering
\includegraphics[width=\textwidth]{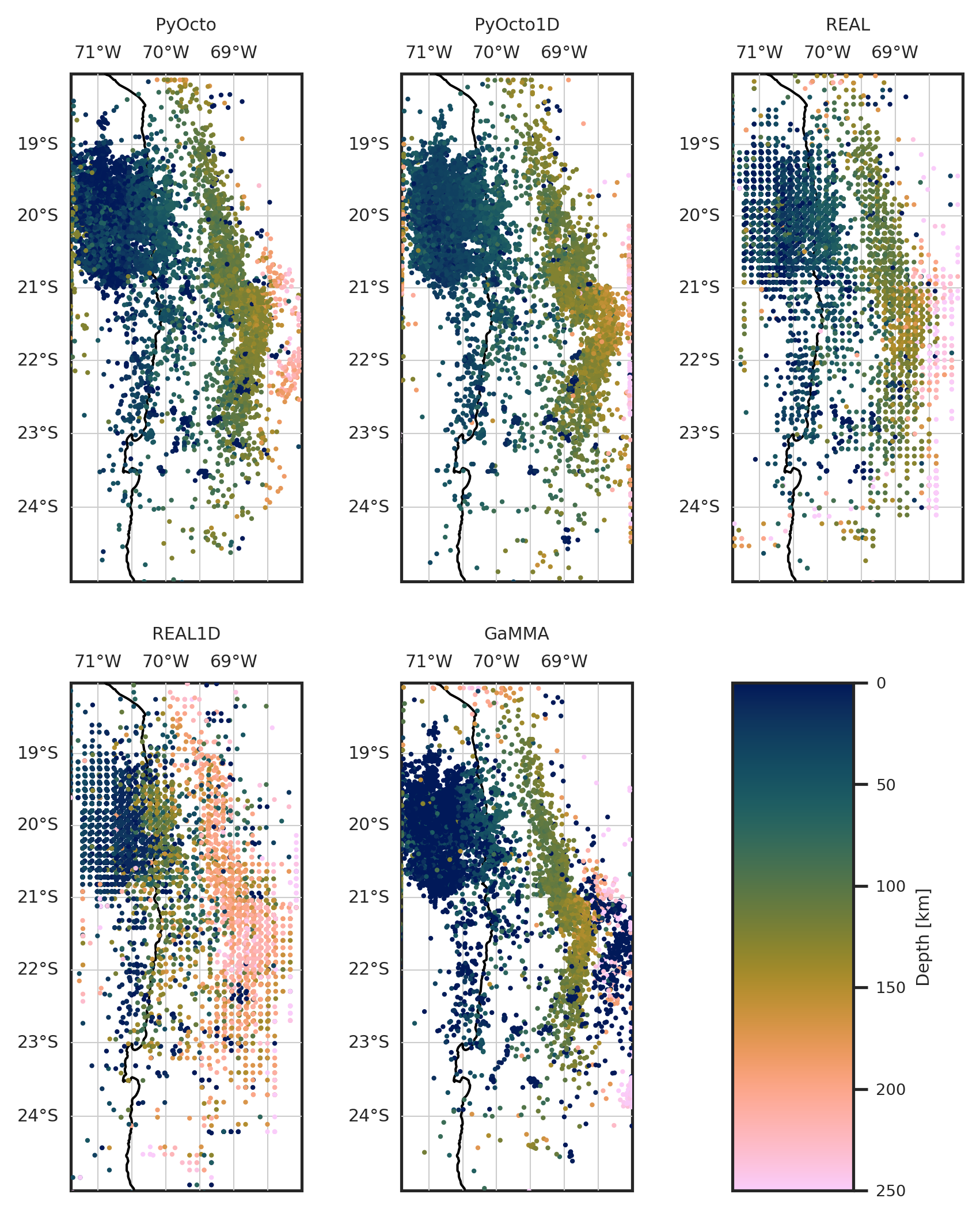}
\caption{Catalogs generated for the Iquique sequence (15th March 2014 to 15th April 2014) using different phase associators. We visualize the output locations as provided by the associators. Please note that in a comprehensive workflow, absolute and relative relocation techniques should be used as a refinement step. Cross section plots are shown in Figure~\ref{fig:iquique_sections}.}
\label{fig:iquique_maps}
\end{figure*}

\begin{figure}[ht!]
\centering
\includegraphics{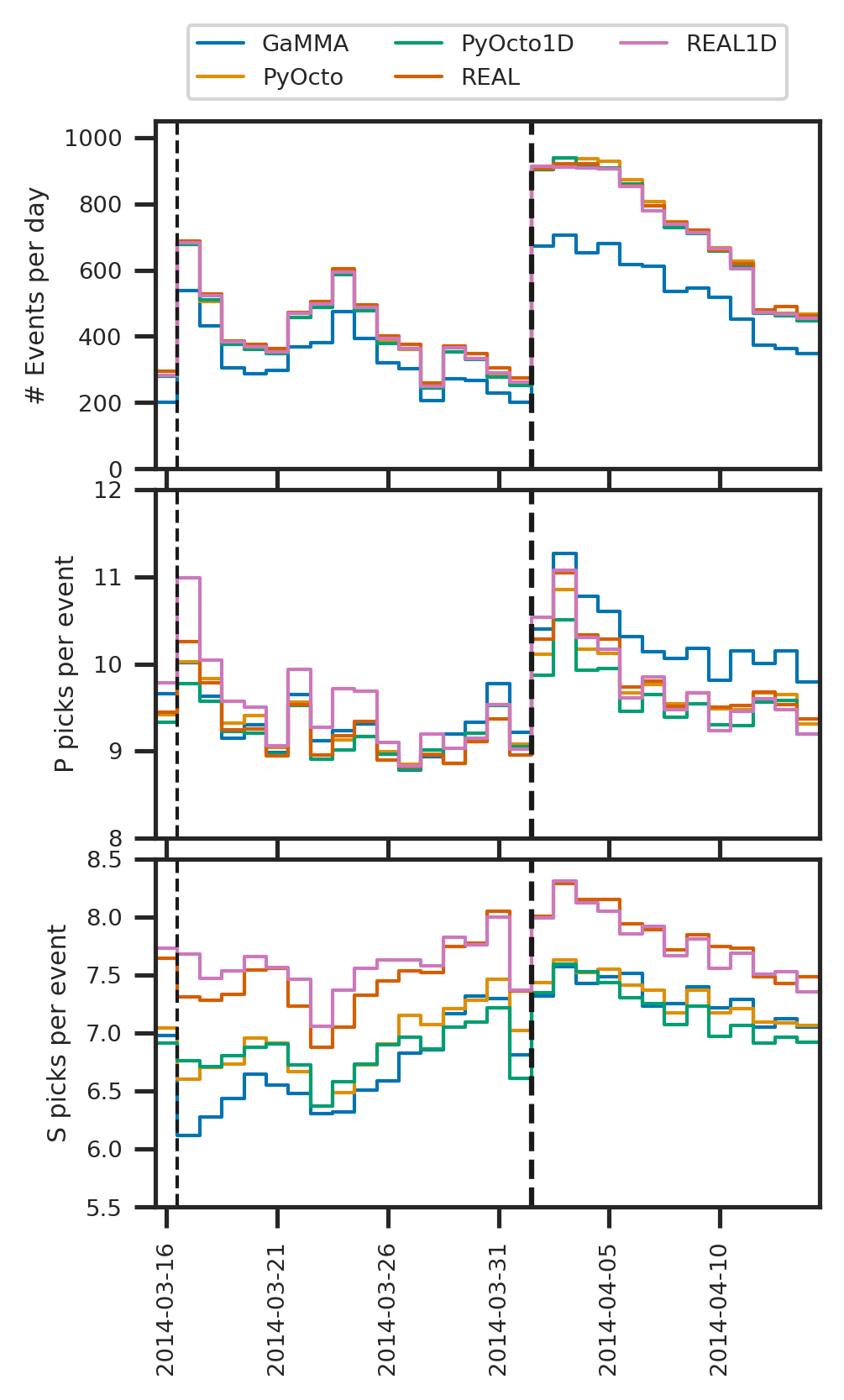}
\caption{Daily earthquake rates, daily number of associated P picks per event, and daily number of associated S picks for the catalogs generated using the different associators. Vertical black lines indicate the times of the largest foreshock and the mainshock.}
\label{fig:iquique_rates}
\end{figure}

\begin{table*}
\caption{Catalog statistics for the Iquique sequence catalog with different associators. The table shows the number of events, picks per event, the fraction of associated picks among all picks, and the total number of picks. We abbreviate \textit{picks per event} as \textit{ppe}. Times refer to average run times per day of data.}
\label{tab:iquique}
\begin{tabular}{lrrrrrrrrr}
\toprule
Associator & Events & Ppe & P ppe & S ppe & Associated & P associated & S associated & Total picks & Time [s] \\
\midrule
GaMMA & 12,718 & 16.92 & 9.90 & 7.02 & 0.34 & 0.31 & 0.39 & 634,647 & 1021 \\
PyOcto & 16,660 & 16.77 & 9.62 & 7.15 & 0.44 & 0.39 & 0.52 & 634,647 & 12 \\
PyOcto1D & 16,362 & 16.56 & 9.49 & 7.06 & 0.43 & 0.39 & 0.50 & 634,647 & 15 \\
REAL & 16,747 & 17.35 & 9.66 & 7.69 & 0.46 & 0.40 & 0.56 & 634,647 & 1487 \\
REAL1D & 16,489 & 17.51 & 9.78 & 7.73 & 0.46 & 0.40 & 0.55 & 634,647 & 1557 \\
\bottomrule
\end{tabular}
\end{table*}

In addition to the synthetic tests, we evaluate the different associators on a real scenario.
For this, we study the 2014 Iquique sequence.
Starting with an 8 month long slow slip transient, the 2014 Iquique sequence contained a magnitude 6.6 foreshock on 16th March and the mainshock on the evening of 1st April \citep{socquet20178, soto2019probing}.
We look at the time between 15th March 2014 and 15th April 2014.
This time span includes the largest foreshock, the mainshock, and the phase of most intensive aftershock activity.
For this study, we use data from the 20 stations in the CX network.
We note that generally more stations from other networks are available in the area.
However, as we do not aim to produce a comprehensive catalog but rather to test the associators, we restrict ourselves to the high-quality CX stations.

Using the CX data, we build a small earthquake detection workflow.
First, we pick P and S arrivals in the continuous waveforms using PhaseNet \citep{zhu2019phasenet} trained on INSTANCE \citep{michelini2021instance} using SeisBench \citep{woollam2022seisbench}.
We use a pick threshold of 0.05 for both P and S waves, i.e., every pick that has a confidence value above 0.05 assigned to it by the deep learning picker is treated as an arrival.
This is intentionally a very low threshold to further stress test the associators.
Second, we pass the picks to each associator to obtain catalogs.
For each associator, we provide picks in daily chunks.
As in our benchmark, we require at least 10 picks and 4 stations with both P and S pick.
We note that this is an extremely simplistic catalog generation workflow that misses essential postprocessing steps, such as absolute and relative relocation or magnitude estimation.
However, it is sufficient to investigate the difference between the associators.

Figure~\ref{fig:iquique_maps} shows the seismicity in the IPOC area, including Northern Chile, as determined with the different associators.
All catalogs clearly show the main features of the seismicity: an intense cluster of events around the Iquique mainshock in the North-West, moderate seismicity along the subducting slab, and a strong band of deeper seismicity.
Table~\ref{tab:iquique} shows statistics for the number of events per catalog, the number of associated picks and the fraction of total picks associated.
Overall, the PyOcto and REAL catalogs are largest, with the catalogs from REAL containing slightly more events.
For both PyOcto and REAL, the catalogs with homogeneous velocity models are slightly larger.
The catalog from GaMMA is about a quarter smaller.
Overall, PyOcto and REAL associated between 43~\% and 46~\% of all picks while GaMMA associated 34~\%.
We note that this does not imply that all remaining picks are incorrect, as many might stem from events that have not been recorded at sufficiently many stations to meet the quality control criteria or even be associated.

Figure~\ref{fig:iquique_rates} shows the daily number of events and the average number of P and S picks per event per day.
Across all days, the number of events is very similar between all variants of REAL and PyOcto, with PyOcto always detecting slightly more events than REAL in the early parts of the aftershock sequence.
GaMMA consistently finds fewer events, with the absolute and relative difference becoming particularly large on days with high seismicity rate.
This indicates that the model is less able to deal with high rates of seismicity.
However, comparing the seismicity rate to the expected Omori decay in activity, it is apparent that all models miss events in the earliest days after the mainshock.
Our results can not distinguish if this is a limitation of the picking model or the association models.

Looking at the average number of picks per event, the only noticeable difference between the associators is that REAL consistently finds about 0.6 S picks more per event.
However, the temporal development of picks per event is interesting.
Overall, the number of P picks per event seems to correlate slightly positively with the total number of events.
For the S picks, the rate of association also follows systematic patterns across all associators, but a correlation with the number of events is not as apparent.
We suggest that the shifts in the number of associated picks are related to the picker performance over time, which is in turn affected by the event distribution.
More large events will cause more impulsive, i.e., easier to detect arrivals.
At the same time, a higher seismicity rate will also cause higher noise levels, making phase detection and picking overall more challenging.

While this study does not focus on the location accuracy of different associators, as we do not perceive this as the main output of the models, we still provide a brief analysis of our findings in the Iquique sequence.
Each method produces a distinct signature of location artifacts (Figures~\ref{fig:iquique_maps} and \ref{fig:iquique_sections}).
GaMMA features a substantial number of shallow detections not present in the other catalogs.
These are primarily mislocations, likely caused by the initialisation of the sources for the expectation-maximization algorithm at the surface.
They occur primarily outside the network.
REAL shows clear gridding artifacts caused by the discretisation of the search grid.
Finer search-grids would reduce this effect, but come at a substantial compute cost, with halving the grid-space leading to roughly 8 times longer run time.
PyOcto shows line-shaped artifacts, however, these are particularly visible with regard to event depth. 
These stripes are caused by failures in the minimization of the EDT loss in the localization procedure.
The artifact is more pronounced for the homogeneous velocity model than the 1D velocity model, likely because the EDT loss is more focused for the 1D model.
Stripes could be reduced or eliminated by increasing the sampling depth in the octotree search for localization.
However, this would lead to increased runtime.
In conclusion, while all associators give a good overview of the general spatial patterns of the seismicity, the locations should only be treated as preliminary estimates.
For accurate location, absolute or relative relocation tools, e.g., NonLinLoc \citep{lomax2000probabilistic} or HypoDD \citep{waldhauser2001hypodd}, should be employed.

We measured average runtimes per day for each associator.
As in the synthetic benchmark, PyOcto was by far the fastest, taking 12~s (homogeneous) / 15~s (1D model).
Gamma took about 17~minutes per day, REAL took 25~minutes (homogeneous) / 26~minutes (1D model).
This means a speed-up factor of 70 to 130 for PyOcto compared to the baselines.
As a reference, loading the waveform data from disk and picking it took around 60 to 90~s per day.
This means that in this scenario, run times for PyOcto association are one order of magnitude below the times for picking, while for the other associators the association largely dominates the total run time.

\section{Conclusion}

In this paper, we introduced PyOcto, a novel seismic phase associator based on space-time partitioning.
We tested PyOcto in two distinct synthetic earthquake scenarios with different numbers of events and different noise levels.
PyOcto consistenly showed detection performance on par or even superior to the state of the art approaches GaMMA and REAL.
At the same time, PyOcto achieves substantial speedups, often with factors above 50.
We furthermore compared the algorithms on the challenging 2014 Iquique sequence.
Here too, PyOcto produces a very complete seismicity catalog.
Similar to the synthetic cases, PyOcto again achieves a speedup of above 70 compared to the other associators, with phase association taking substantially shorter time than the phase picking.
This makes the algorithm future-proof in face of ever-growing seismic networks and potentially more sensisitive, future phase pickers.
PyOcto is available as an open-source tool.

\begin{acknowledgements}
This work has been partially supported by MIAI@Grenoble Alpes (ANR-19-P3IA-0003).
I thank Frederik Tilmann and Marius Isken for insightful discussions that helped improve the algorithm design.
I thank Sophie Giffard-Roisin for her comments that helped improve the manuscript.
\end{acknowledgements}

\section*{Data and code availability}
PyOcto is available at \url{https://github.com/yetinam/pyocto} and Zenodo (publication with DOI in progress).
The code for the benchmark is available in the same repository.
PyOcto can be installed from PyPI using pip.
Waveform data for the CX network (\url{https://doi.org/10.14470/PK615318}) was obtained through the GEOFON FDSN webservice.

\section*{Competing interests}
The author has no competing interests.

% \bibliography{mybibfile}

\end{document}